\documentclass[aps,prl,showpacs,twocolumn,groupedaddress]{revtex4}
\usepackage{graphicx}
\usepackage{psfrag}

\def\bea{\begin{eqnarray}}
\def\eea{\end{eqnarray}}
\def\ba{\begin{array}}
\def\ea{\end{array}}

\begin{document}

\title{Random bearings and their stability}
\author{ R. Mahmoodi Baram}
\email[]{reza@ica1.uni-stuttgart.de}
\author{Hans J. Herrmann}
\email[]{hans@ica1.uni-stuttgart.de}
\affiliation{Institute for Computational Physics, University of
Stuttgart
Pfaffenwaldring 27, 70569 Stuttgart, Germany}

\date{\today}
\begin{abstract}
Self-similar space-filling bearings have been proposed some time ago as models for the motion of tectonic plates and appearance of seismic gaps. These models have two features which, however, seem unrealistic, namely, high symmetry in the arrangement of the particles, and lack of a lower cutoff in the size of the particles. In this work, an algorithm for generating random bearings in both two and three dimensions is presented. Introducing a lower cutoff for the sizes of the particles, the instabilities of the bearing under an external force such as gravity are studied. 
\end{abstract}
\maketitle

\begin{figure}
  \begin{center}
    \includegraphics[width=0.43\textwidth]{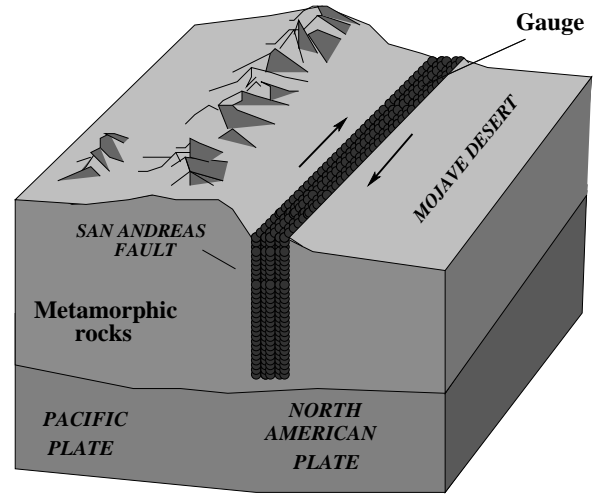}
    \caption{San Andreas fault and nearby geological structure. Different tectonic plates move relatively.}
    \label{fig:tectonic}
  \end{center}
\end{figure}

The term seismic gap refers to any region along an active geological plate boundary that has not experienced a large thrust or strike-up earthquake for more than 30 years \cite{McCann}. Plate tectonic theory uses the concept of seismic gaps to provide very rough estimates on the location and magnitude likeliness of earthquakes. The tectonic plates usually tend to move relative to each other due to the earth's internal convection, but the large friction between the boundaries hinders a continuous sliding which would be several centimeters per year and leads to accumulation of stress over the course of time. Beyond a critical point the accumulated stress is released resulting in big shocks and large relative motions of the plates of up to $20m$. Figure \ref{fig:tectonic} demonstrates San Andreas fault and nearby geological structure and how different tectonic plates move relatively.   

In some faults, like that of San Andreas, tectonic plates have been moving for a long time (thousands of years) without any significant earthquake or production of heat as it is expected for the processes involving rubbing rough surfaces. The understanding of these seismic gaps is one of the big challenges in geophysics. Space-filling bearings were introduced more than a decade ago for the first time as a simplified model for explaining this phenomenon\cite{hans-prl,hans-gen}. In this model, it is assumed that the space between the tectonic plates is filled with more or less round particles which, as the plates move, may roll on each other resulting in the spontaneous formation of local bearings and reducing the amount of friction and dissipation of energy. The spontaneous formation of bearings has been evidenced to be possible in Molecular Dynamics simulations of shear bands \cite{shear-band}, supporting the model. Using techniques based on conformal mapping, originally self-similar space-filling bearings in two dimensions were proposed. They are stripes completely filled with an infinite number of discs of different sizes. Following the same line, three dimensional packing of spheres has been recently constructed. Although the dynamics of bearings in three dimensions is more complicated than in two dimensions due to higher degrees of freedom, it has been shown that the necessary and sufficient condition for a packing of spheres to be a bearing is to be bi-chromatic \cite{3d-bearings}. In other words, only two colors are needed for coloring all spheres in such a way that no spheres of the same color touch each other. 

There are two major criticisms which have been raised about such bearings. First, due to the nature of the construction algorithm the location of the particles are very specific which makes such bearings very unlikely to occur in nature. Second, the space is completely filled with particles of different sizes down to infinitely fine grains, whereas in the reality there exists always a minimum size of the particles. The present work is focused on constructing random bearings and studying the effect of cutoffs for the size of the particles on the stability of the system. In the following, an algorithm for constructing random bearings in both two and three dimensions is presented. In the construction procedure, the formation of odd loops is avoided by imposing the bi-chromatic condition. Next, the instability of the configurations as a consequence of setting a cutoff is discussed and calculations for the dissipation of energy in a system with rotating particles under gravity are presented. One can see, that the energy dissipation decreases as the cutoff is reduced. Finally a discussion of the results and the conclusion is given. 

\begin{figure}
  \begin{center}
    \includegraphics[width=0.16\textwidth]{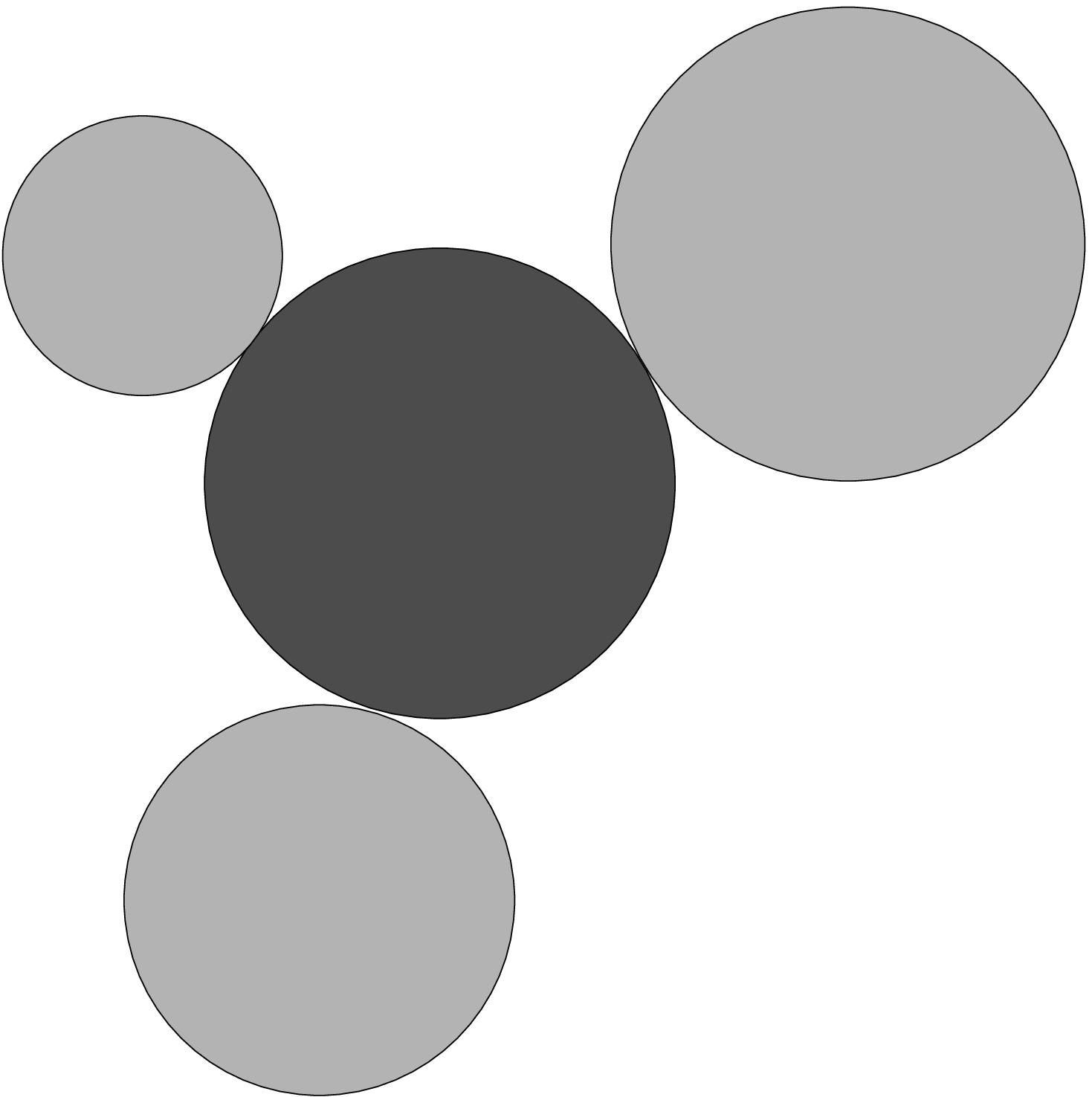}
    \hspace{0.5cm}
    \includegraphics[width=0.16\textwidth]{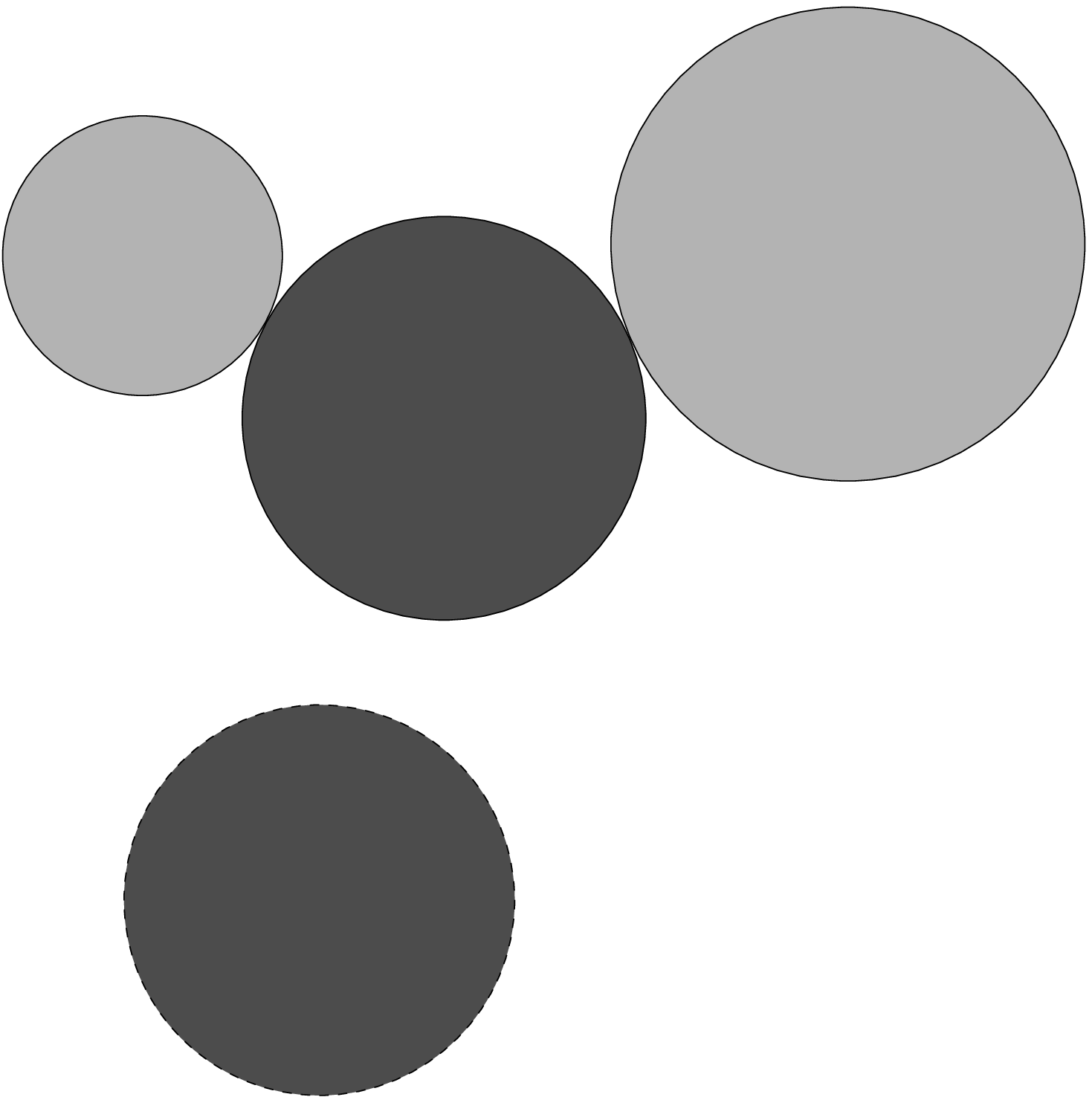}
    \caption{The method for construction of a bi-chromatic packing: If all three discs are of the same color (left image) a disc is inserted with opposite color touching all three, otherwise its size is reduced by a factor $\alpha$ so that it only touches the two that have the same color (right image). } \label{fig:colouring}
  \end{center}
\end{figure} 

The algorithm can be divided in two parts. First we construct a general random packing of discs or spheres. Second we impose the bi-chromatic condition which implicitly guarantees the packing to contain no odd loops and, therefore, be a bearing \cite{3d-bearings}. Initially, some discs within a given range of sizes are randomly distributed in space without touching each other. The filling procedure is continued from then on by inserting always the biggest possible disc into the system without overlapping with any existing disc. This is the most efficient way of filling the space starting from a given initial configuration of discs. This becomes more obvious as the local configurations are observed to be close to that of classic Apollonian which is the most efficient known way of packing discs. For finding the biggest hole where a new disc can be inserted, an arbitrary disc $A$ from the current configuration is chosen and all possible neighboring pairs are examined between which a disc can be inserted touching all three without overlapping any other disc in the system. In this way, the locally biggest disc is found and set as the candidate to be inserted next into the system. This disc will touch the initially chosen disc $A$ and two others, namely, $B$ and $C$. To find the final candidate we check whether the current candidate is also the biggest for $B$ and $C$. In other words, discs $B$ and $C$ are examined as was disc $A$. If a bigger disc is found the candidate for being inserted next is updated. Continuing this search, the biggest disc can be finally found and inserted. More discs are packed into the system by repeating the same procedure over and over. One notices that as the density increases, different regions of the packing become independent. Therefore one doesn't need to look for the globally biggest disc each time and the search can be stopped after a few iterations.

\begin{figure*}[ht]
  \begin{center}
    \includegraphics[width=0.8\textwidth]{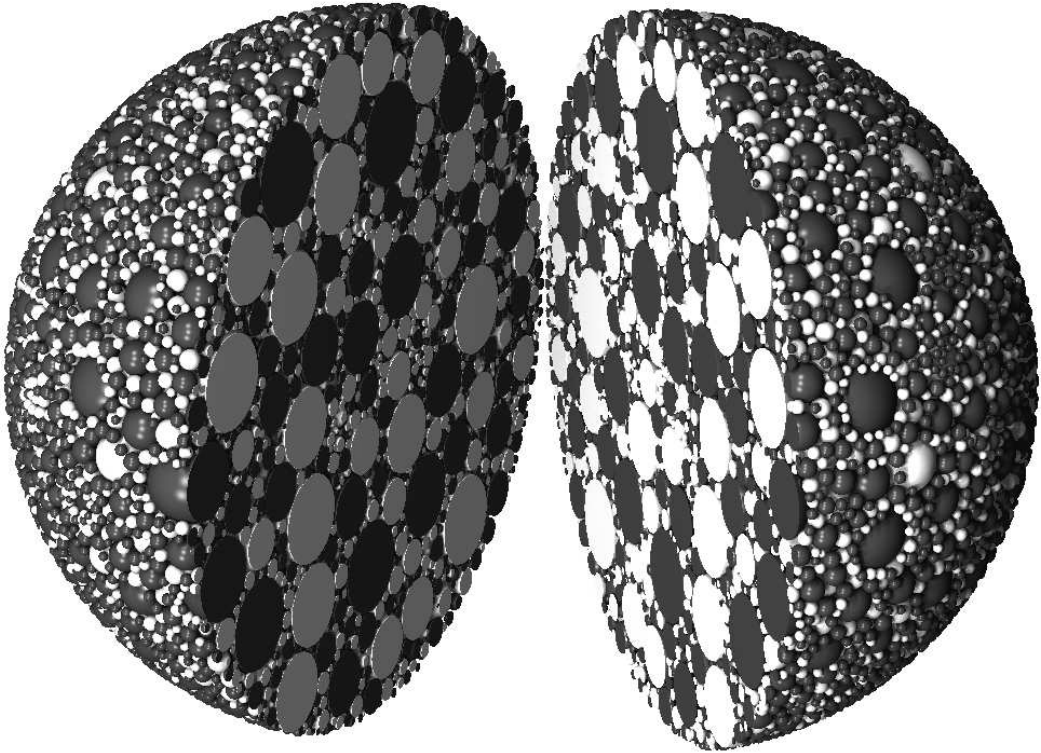}
    \caption{Three dimensional random bearing. No two spheres of the same color touch each other.}
  \label{fig:3d_bearing}
  \end{center}
\end{figure*}

So far no considerations have been made for the packing to act as a bearing. As the consequence, there will be many odd loops in the packing which will hinder any frictionless rotation of the discs. The bearing condition, however, can be easily implemented into the algorithm. The system is initialized as before except that the initial discs are also assigned randomly with two colors. A new disc, gets a color such that it doesn't touch any disc with the same color. This is only possible if all three touching discs have the same color. In other cases, where only two of the discs have the same colors, the radius of the inserted disc is reduced by a factor $\alpha$ with respect to the size which would make it touch to all three:
\bea
r=\alpha r_0,
\label{eq:bearing-factor}
\eea
where $r_0$ is the size of the biggest possible and $r$ is the size of inserted disc as shown in figure \ref{fig:colouring}. For $\alpha=1$ a random packing is obtained which is not a bearing. Therefore, for any $\alpha$ less than unity one obtains a bearing. 

\begin{figure}[t]
  \begin{center}
    \includegraphics[width=0.45\textwidth]{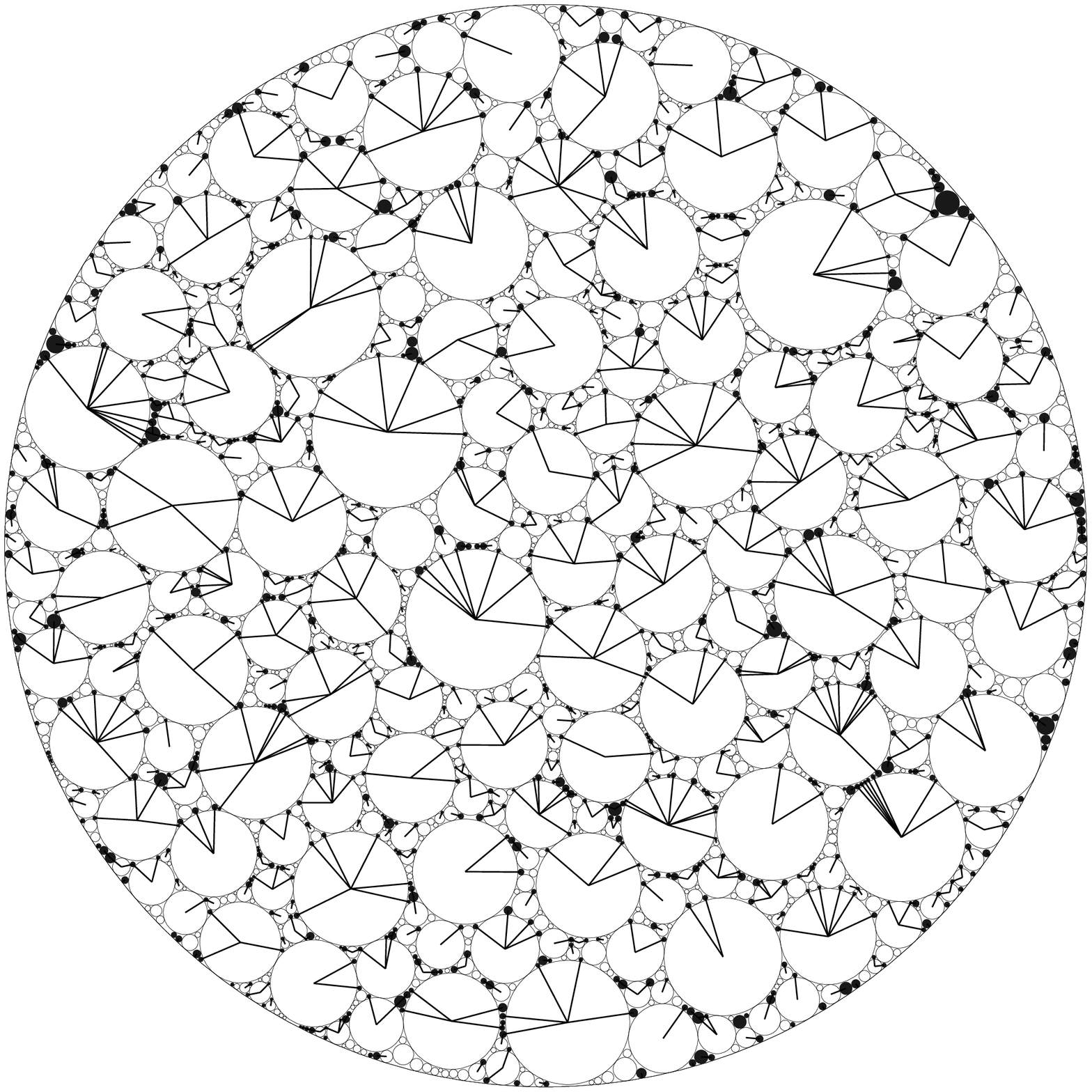}
    \caption{Two dimensional random bearing with $\alpha=0.6$. Applying gravity, some particles move and form frustrated contacts. These are shown as black discs. Solid lines show the frustrated contact.}
    \label{fig:bearing2d}
  \end{center}
\end{figure}

Similarly, random packings and bearings can be obtained in three dimensions with this method. The difference to two dimensions is that each new sphere is inserted touching {\em four} spheres. To construct a bearing three situations should be considered, that is, among the four spheres one is of one color and three of the other color, two are of one color and two of the other or all three have the same color. Figure \ref{fig:3d_bearing} shows a resulting bearing in three dimensions for
$\alpha=0.6$.

From both the computational point of view and that of what happens in reality, a smallest particle size  must inevitably exist. The main consequence of this cutoff $\varepsilon$ are unfilled spaces which may cause instabilities in the system under external forces. In other words, the particles may no longer be fixed in their positions, causing changes in the configuration. As we will see, this plays an important role in the dynamics of the bearing. Here in particular we will study the effect of gravity on the system. In a random bearing with a cutoff, the particles which are not supported from below will be displaced by gravity, resulting eventually in the formation of odd loops in the system. In an odd loop at least one frustrated contact will form as the particles are forced to rotate. These are sources for local dissipation of energy and the system will not act as perfect bearing anymore. 

The stability of the system depends on how loose it is before applying the gravity. Here, we make an estimate for the total dissipated energy in the system. Assuming that the friction acting between two rubbing surfaces follows Coulomb's law, at a frustrated contact, the amount of energy dissipated per unit time is, 
\bea
{\cal E}_{dis}=\mu N v_{rel},
\label{eq:dissipation}
\eea
where $\mu$ is the Coulomb friction coefficient, $v_{rel}$ is the relative velocity of the surfaces of the particles at the frustrated contact, and $N$ is the normal force acting between them. As can be easily verified, the normal force $N$ is proportional to the weight of the dislocated particle. The proportionality factor is a function of the angles between normal forces at the contacts of a particle and the gravity direction. In both two and three dimensions, we assume for all frustrated contacts a typical value for this factor. It should be noted that in two dimensions the relative tangential contact velocity is exactly the same for all contacts, zero for unfrustrated and non-zero for frustrated ones, since all touching pairs of discs can rotate either in the same or in opposite direction. Therefore, we can describe the total dissipation of energy as proportional to the total mass of dislocated particles that produce frustrated loops:
\bea
{\cal E}_{total}\sim {\cal M},
\label{eq:friction}
\eea
which we will consider as the measure for the deviation from a perfect bearing.

To check the effect of gravity on the system, we use a semi-dynamics which is an extention of the one used by Manna et al \cite{manna} to simulate discs under gravity. The particles which don't have enough contacts (at least two in two dimensions and three in three dimensions) to carry their weight will either fall freely or role on one another. A particle is fixed if the line starting at the center of the particle and going in direction of gravity cuts at least one line (triangle) made by connecting two (three) contacts in two (three) dimensions. The process of falling and rolling is performed on all particles one at a time while others are held fixed. Those particles which are in a lower position are treated first and the upper ones later. The programm goes through the list of particles several times and lets them fall and roll until no particle moves further. In this way, the system reaches the final state from which $\cal M$ the total mass of particles forming frustrated contact can be calculated.
\begin{figure}[t]
  \begin{center}
    \includegraphics[width=0.42\textwidth]{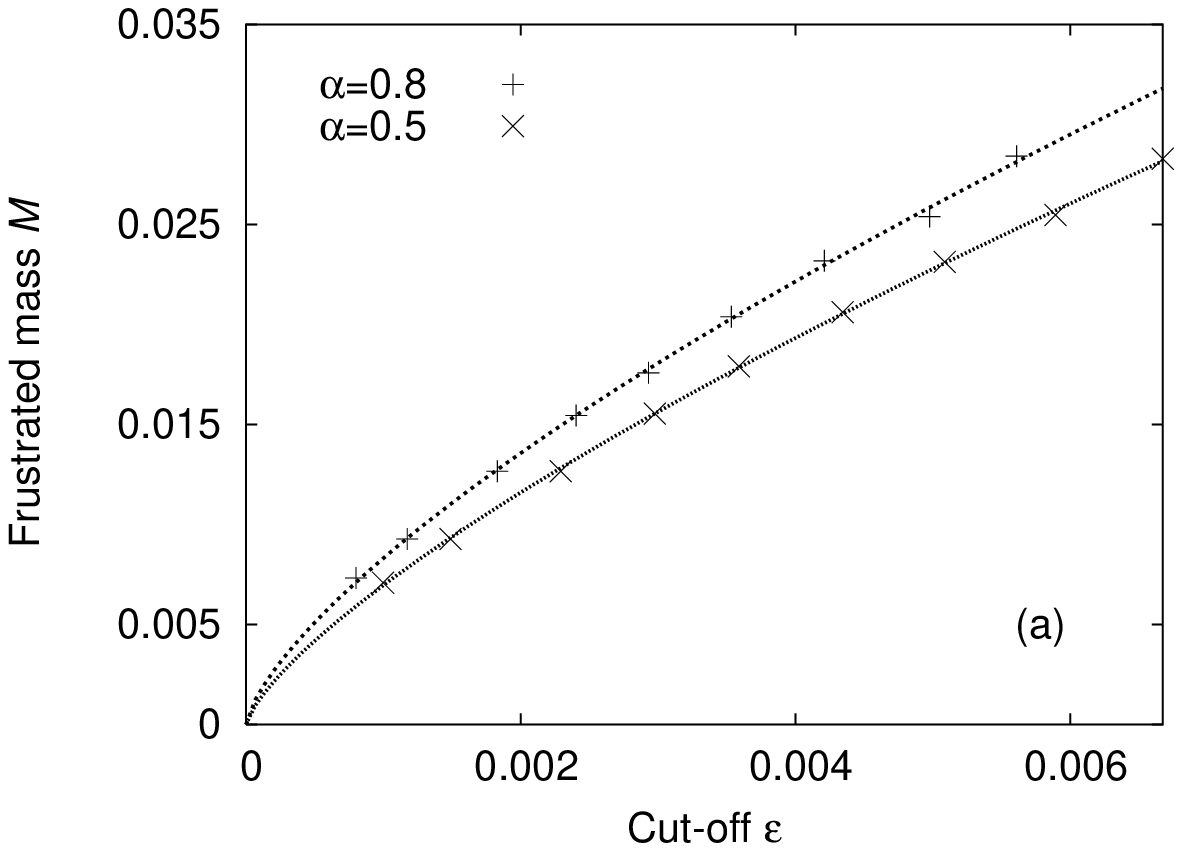}
    \includegraphics[width=0.42\textwidth]{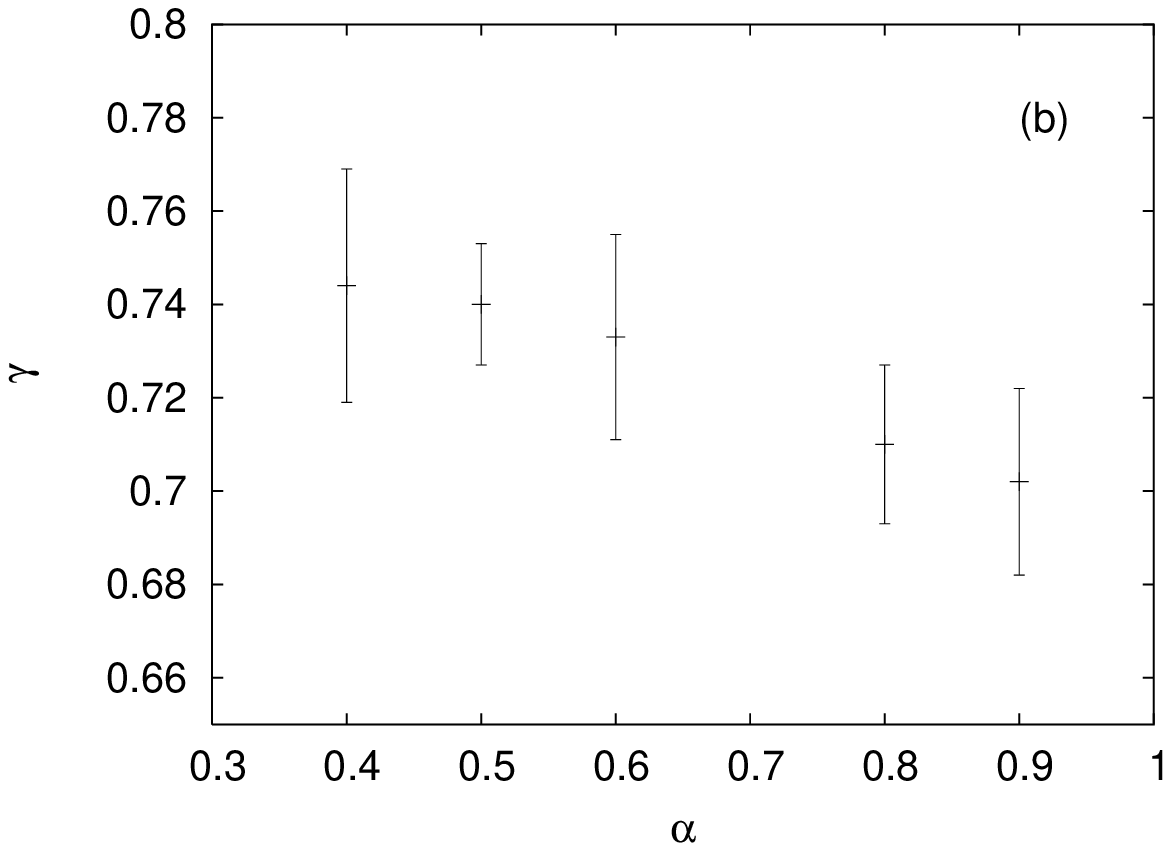}
    \caption{(a) Frustrated mass $\cal M$ as function of the cutoff $\varepsilon$ for two dimensional bearings for $\alpha = 0.5$ and $0.8$. Lines are different power law fits, with exponent $\gamma$= $0.74$ and $0.71$ correspondingly. (b) Exponent $\gamma$ as function of $\alpha$.}
    \label{fig:gravity}
  \end{center}
\end{figure}

Here, we present the calculation for a two-dimensional system. Figure \ref{fig:bearing2d} shows a two dimensional random bearing. Applying gravity, some particles move and form frustrated contacts. These are shown as black discs. Solid lines show the frustrated contacts. The total frustrated mass $\cal M$ is computed as function of the cutoff $\varepsilon$ for different configurations. The result is shown in Fig. \ref{fig:gravity}(a) for two values of $\alpha$. The data points are fitted best by power law functions ${\cal M}\sim \varepsilon^\gamma$. The results indicate that the system approaches the state of complete stability, that is ${\cal M}=0$, as $\varepsilon\rightarrow 0$.

Figure \ref{fig:gravity}(b) shows the calculated exponent $\gamma$ as function of $\alpha$. It can be seen that the exponent $\gamma$ is more or less independent of $\alpha$ having the value approximately $0.72$. In other words, the way in which the packings are constructed does not play an important role in the obtained results. It should be stressed that the fractal dimension of the packings turns out to be also the same for all values of $\alpha$ within the computational error. 

We propose to study experimentally the energy dissipation of a polydisperse mixture of circular discs in a Couette cell as used by Veje et al. \cite{veje} changing the cut-off of the size distribution.

All space-filling bearings, which have been studied in the past, were highly organized arrangements of particles and there was no lower cutoff on the size of the particles in such bearings. These were two main drawbacks in modeling natural phenomena, like tectonic plate motion. Here, an algorithm has been presented for producing space-filling bearings in which the particles do not follow any regular pattern.  We also investigated the stability of bearings with a finite cutoff under gravity and showed that as the system has less porosity less energy is dissipated. The energy dissipation rate follows a power law behavior with respect to the cut-off on the size of the particles.

We would like to thank M. Strau{\ss}  and M. Wackenhut for useful discussions.

\end{document}